# An Exploratory Study of the Relationship between SATD and Other Software Development Activities

Shima Esfandiari
Department of Computer Engineering
Shiraz University
Shiraz, Iran
shima.esfandiari@hafez.shirazu.ac.ir

Ashkan Sami
Department of Computer Engineering
Shiraz University
Shiraz, Iran
sami@cse.shirazu.ac.ir

*Abstract*— Technical Debt is a common issue that arises when short-term gains are prioritized over long-term costs, leading to a degradation in the quality of the code. Self-Admitted Technical Debt (SATD) is a specific type of Technical Debt that involves documenting code to remind developers of its debt. Previous research has explored various aspects of SATD, including detection methods, distribution, and its impact on software quality. To better understand SATD, one comprehension technique is to examine its co-occurrence with other activities, such as refactoring and bug fixing. This study investigates the relationship between removing and adding SATD and activities such as refactoring, bug fixing, adding new features, and testing. To do so, we analyzed 77 open-source Java projects using TODO/FIXME/XXX removal or addition in inline comments as indicators of SATD. We examined the co-occurrence of SATD with each activity in each project through chi-square and odds ratio evaluations. Our results show that SATD removal occurs simultaneously with refactoring in 95% of projects, while its addition occurs in 89% of projects. Furthermore, we found that three types of refactoring - "move class", "remove method", and "move attribute" - occur more frequently in the presence of SATD. However, their distribution is similar in projects with and without SATD.

## I. INTRODUCTION

Technical debt (TD) is a crucial concept in computer software engineering, defined as "*not quite right code which we postpone making it right*" [1]. TD occurs when developers take shortcuts to save time, but it can make the system harder and more expensive to change later on. Non-technical stakeholders may prioritize short-term gains over long-term costs, exchanging short-term profits for long-term costs [2]. To manage TD, it is essential to be aware of its existence and to mark it. One approach is to place phrases such as TODO or FIXME at the top of the code, which notifies developers of the existence of TD. This idea was first introduced in [3] and is known as Self-Admitted Technical Debt (SATD).

Research related to SATD can be categorized into three groups, according to [4]: detection, comprehension, and repayment. The detection group focuses on introducing approaches, methods, and tools for identifying technical debts. Researchers in the comprehension group study the life cycle of TD and how it relates to other software activities. Finally, the repayment group explores ways to eliminate or reduce TD.

Several studies [5]–[8] in the comprehension category have looked at the TD removal and other software development activities. For instance, [5] analyzed 5 Java open-source projects and found that SATD removal often occurs during bug fixing and new feature implementation, with only a few instances co-occurring with refactoring actions. In contrast, [6] and [7] found a significant relationship between SATD and refactoring. Specifically, [6] measured the contribution ratio of different types of refactoring to SATD removal and identified patterns for its elimination.

SATD can persist in code for extended periods, sometimes for over a decade, and can eventually lead to a decline in code quality [3]. In contrast, refactoring can enhance code quality and can be viewed as a way of metaphorically paying off technical debt [9]. Both SATD and refactoring are crucial indicators for measuring code quality. Refactoring involves "*improving the structural design of a software system without changing its external behavior*" [10].

Previous studies have extensively investigated the simultaneous occurrence of removing Self-Admitted Technical Debt (SATD) with other software development activities, although the number of programming language projects examined was limited. In our study, we conducted an analysis of 77 Java projects and, to the best of our knowledge, examined the simultaneous occurrence of adding SATD with other activities for programming language projects for the first time. Earlier studies found a strong connection between removing SATD and refactoring. There was a significant correlation in over 90% of projects (P-Value < 0.05), and 50% of projects had bug fixes.

The primary objective is to address three questions.

**RQ1: What is the correlation between Self-Admitted Technical Debt (SATD) and other software development activities?** To investigate this, the study examines the co-occurrence of SATD with other activities. The results reveal SATD is present alongside refactoring in 85% of the projects,

occurring over 40% of the time. SATD is found to co-occur with bug fixing in half of the projects.

**RQ2: Does refactoring involve the removal or addition of SATD?** The evaluation employs the chi-square test and odds ratio. The findings show that the co-occurrence of SATD removal (addition) and refactoring is statistically significant in 95% (89%) of the projects. The odds ratio test for all projects also shows values above 1. The chi-square test conducted for the co-occurrence of SATD with bug fixing shows significance in half of the projects, with the odds ratio exceeding 1 in almost all projects.

**RQ3: Does the distribution of refactoring types differ when SATD is present or absent?** The study calculates the cosine similarity of refactoring types per project, both in the presence of SATD and in general. In 89% of the projects, the similarity measure exceeds 0.72. The average correlation between types in the presence of SATD and the general case is 0.84. Consequently, the type of refactoring does not follow a specific pattern when SATD is present.

## II. RELATED WORK

Numerous studies have been conducted on code evolution, how to manage it, and how to identify technical debt. However, none of these studies consider the content of comments in the source code. It wasn't until 2014 that researchers began to pay attention to this aspect. Technical debt can be defined at different levels, as demonstrated in [9], which introduces 10 types of technical debt, including TD at the code, design, and documentation levels. One of these types is SATD, which is embedded in comments. A study by [10] indicates that SATD leads to fewer defects in the future than other types of technical debt. However, making changes to SATD can be more difficult and complex.

Various methods are available to detect SATD, ranging from pattern recognition in comments [11] to more advanced techniques such as natural language processing [12], [13], and machine learning [14]. Once TDs have been identified, developers can use tools to resolve them and improve code quality. SATDBailiff, provided by [15], is an example of a tool designed for detecting Self-Admitted Technical Debt (SATD) based on open Java source projects. However, a more recent automated identifier [16] has been developed that integrates four distinct sources: source code comments, commit messages, pull requests, and issue tracking systems.

In [12], the authors employed 10 open-source projects to demonstrate that words representing low-quality code or future completion needs can be valuable in addition to keywords and fixed phrases. They found that with only 23% of the comments, 90% of the best-ranking performance could be achieved, and with 5 to 9% of the comments, 80% accuracy could be obtained.

Similarly, in [13], the authors presented a method for accurate TD detection. They suggested that simultaneous static code analysis and natural language processing could increase accuracy by identifying written notes from developers that are relevant to TD. An open-source tool/plugin was developed based on this method to detect TD.

Understanding SATD is crucial for the effective management of technical debt. One example of this is demonstrated in [17], which employs big data analysis on 159 projects to answer questions related to the propagation and evolution of SATD. These questions include whether the number of SATDs increases or decreases over time, how long they remain in the system, who fixes them, and how often they are fixed.

Other studies focus on the last stage of the TD life cycle when it is time to delete it. For instance, [8] investigates the number of TD removed accidentally. The findings indicate that 25% to 60% of TDs are deleted during class or function removal, while 33% to 63% are eliminated during changes in function. The remaining TDs are eliminated without a related change, which could be due to the acceptance of TDs by developers.

Technical debt poses a significant challenge for researchers, who aim to detect TDs more accurately and gain a better understanding of them to improve code quality and reduce future costs. Many studies have examined the life cycle of TDs, particularly during the elimination stage. For instance, [18] conducted an analysis of 102 systems that depend on databases and found that the primary causes for the insertion of data-access Self-Admitted Technical Debts (SATDs) are bug fixing and refactoring activities. However, it is important to note that this study was limited to analyzing database systems only.

The study that is most similar to ours is [19], which also examines the co-occurrence of removing Self-Admitted Technical Debts (SATDs) with refactoring in our dataset. However, our study goes beyond that and considers other software development activities in both the removal and insertion of SATDs.

This study falls under the comprehension category and, similar to previous works, focuses on the removal of SATDs while also considering their insertion. However, what distinguishes this research is the use of a substantially larger dataset of programming language projects compared to previous studies. For example, [6] examined the co-occurrence of SATD and refactoring in only four projects, while [5] and [12] investigated this issue in five and ten projects, respectively. In contrast, our study analyzes a much larger dataset, allowing for a more comprehensive understanding of SATD dynamics. Additionally, our study provides a deeper insight into the specific types of refactoring that co-occur with Self-Admitted Technical Debts (SATDs). This allows us to gain a better understanding of the relationship between SATDs and different refactoring activities, providing more nuanced insights into software development practices.

## III. DATA SET AND THE PROPOSED METHOD

To conduct this study, we used data from SmartSHARK [20]. SmartSHARK provides information on the evolution of software projects extracted from various sources, including GitHub, Apache Jira, and Travis CI. The data was integrated into a database, which included 77 Java projects that adhere to high standards for development processes. The selected projects cover a wide range of applications, such as building systems, web development, and big data processing tools.

Furthermore, the chosen projects are of medium size and can be found in the Appendix.

The database used in this study categorizes commits into 12 types, which were classified into five general categories: bug fixing, documenting, refactoring, adding new features, and testing. Additionally, there are 35 groups of refactoring types. SATD is categorized under documentation, and its presence is indicated by TODO and FIXME comments added or subtracted from the source code. Therefore, the co-occurrence of each category (add and remove SATD) with refactoring was investigated separately. Whenever SATD was present along with another tag on a commit, it indicated the co-occurrence of the two.

To investigate the relationship between SATD and other software development activities, this study proposes a method that counts the number of times SATD co-occurs with another category. The most frequently occurring activity may be statistically related to technical debt and statistical tests are performed to verify this relationship. The results and evaluation section of this study provides answers to the research questions.

As illustrated in Fig. 1 our experiment design involves a series of steps. Firstly, we utilize the SmartShark dataset as our input. Next, we categorize software development activities into five main elements, with an additional element considered as well. This categorization enables us to determine the distribution of these activities. Furthermore, we delve into the co-occurrence of self-admitted technical debt (SATD) removal/insertion alongside other software development activities. Additionally, we specifically examine the co-occurrence of SATD removal and other software development activities at the project level, using statistical tests such as chi-square and odds ratio analysis. Lastly, we analyze the distribution of various refactoring types in the presence of SATD and in a general context. The primary objective of this comprehensive experiment design is to gain valuable insights into the relationships and patterns existing between SATD and other software development activities.

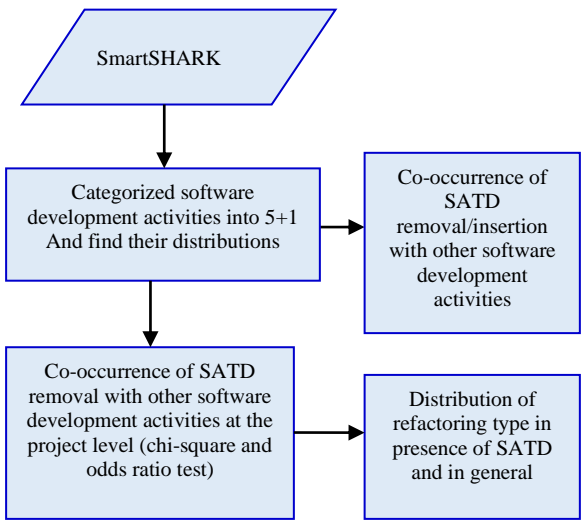

Fig. 1. Summary of our experiment design

The aim of the paper is to quantify the ratio of the occurrence of SATD with other software development activities using the odds ratio. This involves calculating a 2×2 contingency table like the Table 1, which calculates the co-occurrence of SATD and refactoring.

Table 1. An example of a contingency table for calculating the odds ratio

|  | SATD | ~SATD |
|---|---|---|
| Refactoring | A | B |
| ~Refactoring | C | D |

Mathematically, the odds ratio is expressed as AD/BC. A value greater than 1 shows a positive association between the variables, suggesting that the event is more likely to occur in the exposed group. Conversely, a value less than 1 shows a negative association, implying that the event is less likely to occur in the exposed group. An odds ratio of exactly 1 suggests no association between the variables.

The paper also uses the chi-square test to see if there is a strong connection between two categorical variables. This test compares observed and expected frequencies to see if variables are independent. The test calculates a chi-square ($\chi^2$) test statistic, which follows a chi-square distribution.

To perform the chi-square test, a null hypothesis is performed that assumes no association between SATD and any of the five categories. The relevant variables are then calculated using the formula:

$$\chi^2 = \sum \frac{(O_i - E_i)^2}{O_i} \quad (1)$$

Where Oi and Ei represent the observed and expected values, respectively. The chi-square test calculates a p-value that tells us the chance of getting our data if the variables were not connected. If the p-value is less than 0.05, it shows a significant connection between the variables, and the null hypothesis is rejected.

IV. RESULTS AND EVALUATIONS

Fig. 2 displays the distribution of each category of data set commits. The data label is well-balanced, except for the number of SATD and documents, which is different. This is because SATD is a subset of documentation and cannot be compared separately. Therefore, both categories are ignored in the analysis.

To answer Research Question 1, this study first investigates the general relationship between SATD and other software development activities, followed by a project-oriented analysis. The study counts the number of times each tag occurs with SATD removal or addition in a commit. Fig. 3 and Fig. 4 display the results of this analysis.

Refactoring accounts for only 14% of the tags at the general level, which is lower than other activities. However, it is the most common activity that co-occurs with SATD, as shown in Fig. 3 and Fig. 4. To investigate whether this relationship holds at the project level, Fig. 5 presents the project-level analysis. In response to Research Question 2, the study finds that refactoring occurs both with the addition and deletion of SATD, as observed in Fig. 3 and Fig. 4.

Fig. 5 presents the project-level analysis of the relationship between SATD and other software development activities, where the "Number of SATD" section shows the percentage of SATD in each project. In some projects, the total percentage of adding new features, refactoring, and bug fixing exceeds 50% due to overlapping categories. The study performs chi-Square tests and calculates p-values for four modes of co-occurrence in each project: the presence of both, absence of refactoring with the presence of SATD, absence of SATD with the presence of refactoring, and absence of both. The results indicate that in 95% of the projects, SATD removal is significantly associated with refactoring, with a P-value less than 0.05. Similarly, in 89% of the projects, SATD addition is significantly associated with refactoring, with a p-value less than 0.05. The odds ratio test also shows values above 1 for all projects.

The results of the chi-Square test show that there is a significant co-occurrence between SATD and bug fixing in 34 projects. This finding may be useful for classifying projects into different categories and identifying the underlying causes of this co-occurrence, which could ultimately lead to more effective strategies for managing SATD and improving software quality.

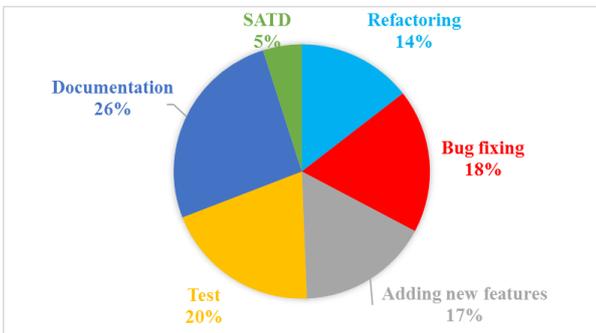

Fig. 2. The distribution of each category of data set commits

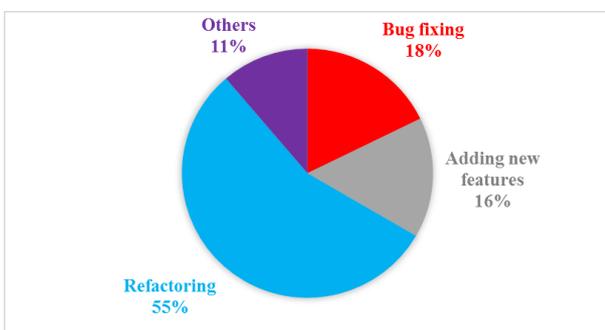

Fig. 3. Co-occurrence of SATD removal with other software development activities

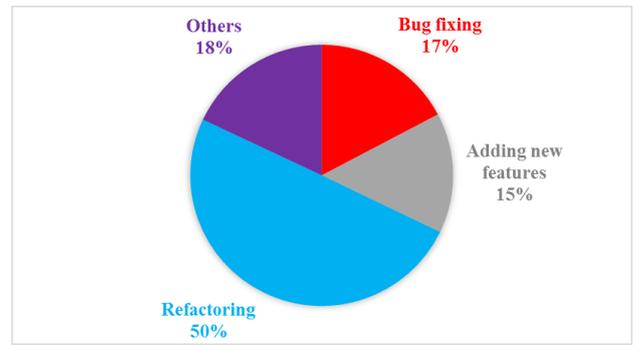

Fig. 4. Co-occurrence of SATD addition with other software development activities

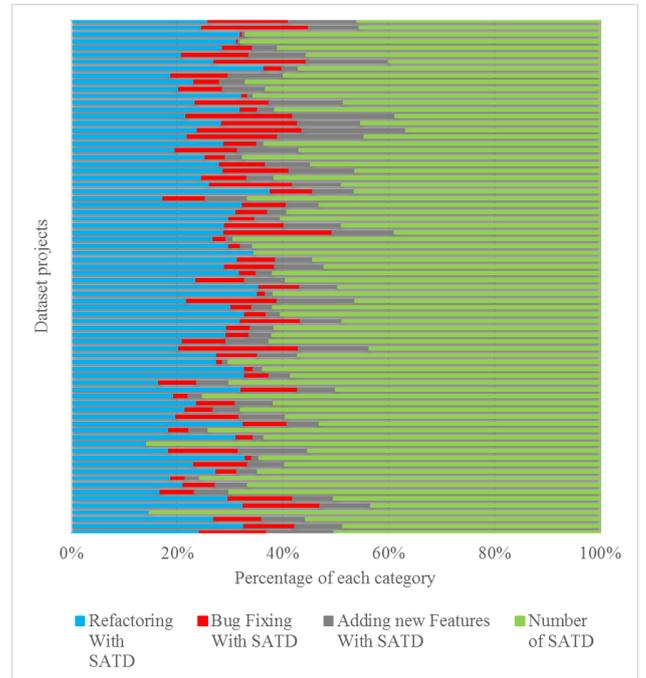

Fig. 5. Co-occurrence of SATD removal with other software development activities at the project level

Recent research has confirmed the close association between refactoring and self-admitted technical debt (SATD). To gain a more comprehensive understanding of this phenomenon, we will explore the relationship between the type of refactoring and SATD. This will enable us to answer Research Question 3.

The study has calculated the frequency of each type of refactoring with and without considering the presence of SATD. The results are presented in Fig. 6, which displays the distribution of the 35 different refactoring types, identified by their respective ID numbers ranging from 1 to 35. The names of these refactoring types are listed in Fig. 7.

For instance, in Fig. 6, the first ID corresponds to the "variable type change" refactoring, the second to the "move class" refactoring, and so on. The horizontal axis represents the ID numbers of the refactoring types, while the vertical axis shows the corresponding count for each type.

The analysis of Fig. 6 revealed that the total number of refactoring types in the presence of SATD shows a different distribution than the general case. While the trend is descending, there are three specific points where the count of refactoring types with SATD reaches its highest. These three points correspond to the refactoring types "move class", "remove method", and "move attribute". However, the overall distribution of the refactoring types remains the same in both cases.

Fig. 7 displays the ratio of each type of refactoring for each project. The horizontal axis shows dataset projects, while the vertical axis shows the ratio of each refactoring type. At the general level, the three most commonly used refactoring types are "change variable type", "move class", and "extract method". However, in the presence of SATD, "move class" is the most frequently used refactoring type.

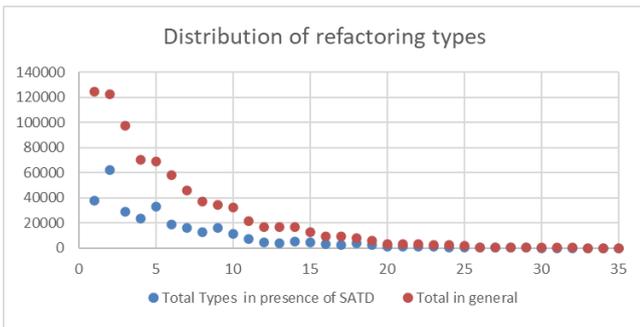

Fig. 6. Distribution of refactoring type in the presence of SATD and in general

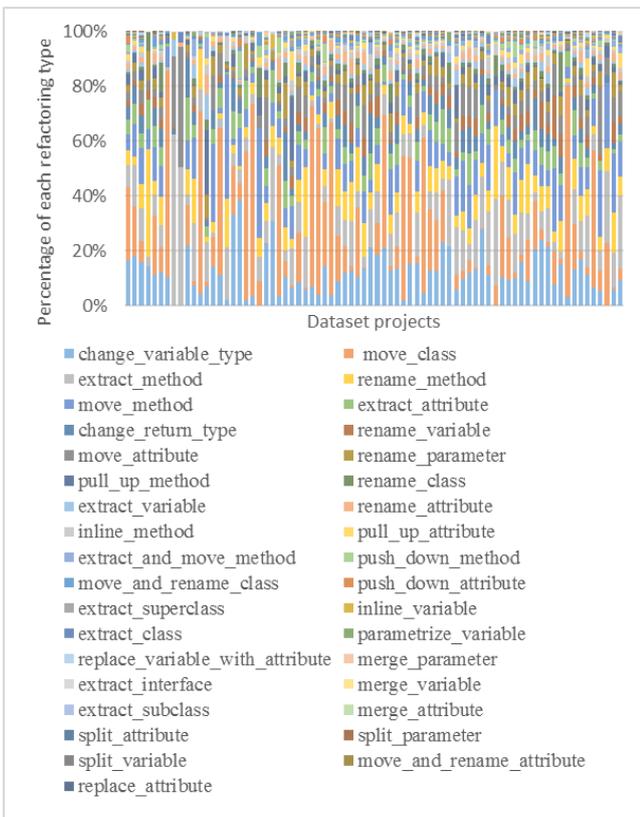

Fig. 7. Distribution of 35 refactoring types for 77 projects

## V. THREATS TO THE VALIDITY

External Threat Validity: Although this paper uses 77 Java projects, which is a larger sample size than previous studies, it may not represent all software projects as all the projects are in the same language and of medium size. Additionally, the paper only addresses the relationship between SATD and other activities and not all types of technical debt.

Internal Threat Validity: The relationship between SATD and refactoring was analyzed by counting the number of co-occurrences and using the Chi-Square test and Odds-Ratio. However, it is possible that both measures may be influenced by a third factor, which may affect the study's internal validity.

## VI. CONCLUSION AND FUTURE WORK

Gaining a deeper understanding of self-admitted technical debt (SATD) facilitates its detection, and by successfully identifying SATD, it becomes possible to proactively prevent its persistence in the source code, thereby enhancing the overall code quality. This paper aims to investigate the relationship between SATD and other software development activities, including refactoring, bug fixing, adding new features, and testing, using a dataset of 77 projects larger than previous studies. The study counts the number of co-occurrences between SATD and other activities at both a general and project-specific level. Given the high frequency of refactoring co-occurring with SATD, statistical tests such as Chi-Square and Odds-Ratio were conducted. The results indicate that in 95% of the projects, SATD removal and refactoring occurred simultaneously, while in 89% of the projects, SATD addition and refactoring occurred together.

These are excellent questions that can be explored in future studies. Understanding why different types of refactoring behave differently in the presence of SATD could provide valuable insights into how to prioritize refactoring efforts. Examining whether the observation exists at the project level could help determine the generalizability of the findings. Investigating why SATD and bug fixing occur together in some projects but not others may reveal important contextual factors that influence software development practices. Categorizing projects and analyzing the differences could provide a deeper understanding of the relationship between SATD and software development activities. Finally, analyzing the sequence of activities could help determine whether certain activities tend to precede or follow SATD and refactoring, which could inform the design of more effective software development processes.

## APPENDIX

Names of projects: zookeeper, zeppelin, XML graphics-batik, xerces2-j, wss4j, tika, tez, systemml, struts, streams, storm, santuario-java, samza, roller, ranger, pig, phoenix, pdfbox, parquet-mr, opennlp, openwebbeans, openjpa, oozie, nutch, nifi, mina- sshd, manifoldcf, mahout, lens, kylin, knox, kafka, jspwiki, jena, jackrabbit, httpcomponents-core, httpcomponents- client, helix, gora, giraph, freemarker, flume, fineract, falcon, eagle, directory-studio, directory-kerby, directory-fortress-core, derby, deltaspike, cxf-fediz, commons-vfs, curator, commons-validator, commons-scxml, commons-rdf, commons-math, commons-lang, commons-jexl, commons-jcs, commons-io, commons-imaging, commons-digester, commons-dbcp, commons-configuration, commons-compress, commons-collections, commons-codec, commons-beanutils, commons-bcel, calcite, cayenne, ant-ivy, archiva, bigtop, Activemq